\documentstyle[preprint,aps,prl,psfig]{revtex}
\tighten
\draft
\begin{document}

\draft

\preprint{31. 7. 2000}
\title{Frequency dependent specific heat of viscous silica}
\author{Peter Scheidler, Walter Kob \footnote{Author to whom correspondence 
should be addressed to. e-mail: Walter.Kob@uni-mainz.de}, Arnulf Latz, 
J\"urgen Horbach, and Kurt Binder}
\address{Institut f\"ur Physik, Johannes Gutenberg-Universit\"at,
Staudinger Weg 7, D--55099 Mainz, Germany}
\maketitle

\begin{abstract}
We apply the Mori-Zwanzig projection operator formalism to obtain
an expression for the frequency dependent specific heat $c(z)$ of a
liquid. By using an exact transformation formula due to Lebowitz {\it et
al.}, we derive a relation between $c(z)$ and $K(t)$, the autocorrelation
function of temperature fluctuations in the microcanonical ensemble.
This connection thus allows to determine $c(z)$ from computer simulations
in equilibrium, i.e. without an external perturbation. By considering the
generalization of $K(t)$ to finite wave-vectors, we derive an expression
to determine the thermal conductivity $\lambda$ from such simulations. We
present the results of extensive computer simulations in which we use the
derived relations to determine $c(z)$ over eight decades in frequency,
as well as $\lambda$. The system investigated is a simple but realistic
model for amorphous silica. We find that at high frequencies the real part
of $c(z)$ has the value of an ideal gas. $c'(\omega)$ increases quickly
at those frequencies which correspond to the vibrational excitations of
the system. At low temperatures $c'(\omega)$ shows a second step. The
frequency at which this step is observed is comparable to the one at which
the $\alpha-$relaxation peak is observed in the intermediate scattering
function. Also the temperature dependence of the location of this second
step is the same as the one of the $\alpha-$peak, thus showing that these
quantities are intimately connected to each other. From $c'(\omega)$ we
estimate the temperature dependence of the vibrational and configurational
part of the specific heat. We find that the static value of $c(z)$
as well as $\lambda$ are in good agreement with experimental data.

\end{abstract}


\section{Introduction}
\label{sec1}

If a glass-forming liquid is cooled, dynamic observables, such as
the viscosity, the diffusion constant or the intermediate scattering
function, show a dramatic slowing down~\cite{books,mct_reviews}. At a
certain temperature, the kinetic glass transition temperature $T_g$,
the typical relaxation time of the system exceeds the time scale of
the experiment and the system falls out of equilibrium, i.e. become a
glass\cite{footnote1}. In contrast to these {\it dynamical} observables,
all {\it static} quantities, such as the volume or the enthalpy, show
upon cooling only a relatively weak temperature dependence for $T>T_g$
and the glass transition temperature is noticed only by a gentle change of
slope in their temperature dependence. This change of slope is reflected
in various susceptibilities, such as the thermal expansion coefficient
or the specific heat $c_V$, as a relatively sudden drop when the glass
transition temperature is reached. The physical picture behind this drop
in, e.g., $c_V$ is that at $T_g$ the structural degrees of freedom fall
out of equilibrium and the specific heat reduces to the one of a solid
having only vibrational degrees of freedom. Hence the height of the drop
is usually used to estimate that part of the specific heat associated with
the configurational degrees of freedom. Thus we see that this type of
experiments can be used to probe the dynamics of the structural degrees
of freedom, although the information obtained is rather indirect.

Since the value of $T_g$ is given by the timescale of the experiment,
it can be changed if the sample is cooled with different cooling
rates~\cite{bruning92,vollmayr96a}. Therefore this dependence opens in
principle the possibility to investigate the temperature dependence of the
configurational part of the specific heat. However, since the relaxation
time of the system changes very rapidly with temperature, it is necessary
to change the cooling rate significantly (several decades) in order to see
a significant shift in $T_g$, which is experimentally rather difficult. An
alternative way to probe the dynamics of the structural degrees of
freedom by means of the specific heat is to measure $c(z)$, the (complex)
frequency dependence of the specific heat of the system {\it in its
equilibrium state}. Seminal work in this direction was done by Nagel and
coworkers \cite{birge85,birge86,dixon88,dixon90,menon96} who proposed an
experimental setup that allowed to measure $c(z)$ also at frequencies
as high as 10$^4$ Hz. Independent work in this direction was also done
by Christensen and others~\cite{christensen85,leyser95,jeong95,beiner96}.

Using their measurements of the frequency dependence of the specific
heat, Jeong and Moon were able to predict cooling and heating
curves in differential scanning calorimetry (DSC) and found good
agreement between their predicted curves and the experimental
ones~\cite{jeong95}. Similar results have been found in molecular
dynamics computer simulations~\cite{scheidler_dipl,scheidler00b}. Thus
these investigations give evidence that the information contained in
the equilibrium quantity $c(z)$ allows one to understand also the
out-of-equilibrium situation that is encountered in DSC experiments around
$T_g$. A short review of this can be found in the paper by Simon and
McKenna~\cite{simon97}.

The theoretical interpretation of the
measurements of $c(z)$ was for some time rather
controversial~\cite{oxtoby86,jackle86,zwanzig88,gotze89,nielsen96,jackle90}
since the origin of the frequency dependence of $c(z)$ was not really
clear. Some of these issues could be clarified by experiments done by
Birge~\cite{birge86} and Menon~\cite{menon96}. A partial discussion of
this dispute is reviewed in Ref.~\cite{simon97}. Very recently Nielsen
put forward a theoretical description of $c(z)$ on purely thermodynamic
arguments~\cite{nielsen99}. In the present paper we will use a statistical
mechanics approach to derive a microscopic expression for $c(z)$ and use
this expression, which is identical to the one derived independently
in Ref.~\cite{nielsen99}, to calculate this quantity from a molecular
dynamics computer simulation of silica. Interestingly enough this
expression has been used before by Grest and Nagel~\cite{grest87} to
calculate $c(z)$ from a computer simulation of a simple liquid. However,
in that paper no derivation was given since it seems to have come out
``from a stroke of genius''~\cite{nagel}. Due to limitations of computer
resources the resulting curves for $c(z)$ were unfortunately rather noisy,
a feature they share with the ones in Ref.~\cite{nielsen99}. In contrast
to this, great care was taken in the present work to obtain reliable data,
which in turn allows us to compare the temperature dependence of $c(z)$
with the one from other dynamical quantities, such as the intermediate
scattering function. Thus this permits us to compare the relaxation
dynamics as probed by $c(z)$ with the one observed by more microscopic
methods, such as neutron or light scattering experiments.

The outline of the paper is as follows: In the next section we will use the
projection operator formalism to derive a connection between $c(z)$ and
microscopic variables. In Sec.~\ref{sec3} we will give the details of the
model and the simulations. In Sec.~\ref{sec4} we will present the results and
end in Sec.~\ref{sec5} with a summary and discussion.

\section{Theory}
\label{sec2}

In this section we derive an expression that relates the frequency
dependent specific heat to the autocorrelation function of kinetic
energy fluctuations. Furthermore we show how the thermal conductivity
can be calculated from the generalization of this correlation function
to finite wave-vectors.

For a dense simple liquid under triple point conditions the only slow
modes are the local densities of the macroscopically conserved quantities,
i.e. the number of particles, the total momentum, and the energy. 
For an $N-$particle system these local densities have the form
\begin{equation}
X({\bf r},t) = \sum_{i=1}^N X_i(t) \delta({\bf r} - {\bf r}_i(t)),
\label{eq1}
\end{equation}
where for $X_i(t) = 1$  we obtain the local density fluctuation $\rho({\bf
r},t)$, for $X_i(t)= p_i^\alpha$ the momentum density in direction
$\alpha \in \{ x,y,z\}$,  and for $X_i(t) = E_i:={\bf p}^2_i/2m +
\frac{1}{2} \sum_{j=1}^N V(|{\bf r}_i - {\bf r}_j|)$ the local energy
density fluctuation. (Here $V({\bf r})$ is the potential energy.) 
In the hydrodynamic limit the equation of motion
for these densities  can be derived by using the exact conservation
laws, some (phenomenological) constitutive equations that relate the
spatial derivatives of the densities to their currents, and thermodynamic
relations between the fluctuations of the densities and thermodynamic
quantities such as temperature,  pressure etc.~\cite{forster83}. This
approach is valid as long as there are no other slow, non-hydrodynamic
processes in the system.  Well known examples, for which constitutive
equations and thermodynamic relations have to be modified by, e.g.,
introducing frequency dependent thermodynamic derivatives, are
second order phase transitions or systems with internal degrees of
freedom~\cite{ferell85}.

As mentioned in the Introduction, the dramatic slowing down of the
dynamics of liquids upon cooling is observed on all length scales.
But contrary to second order phase transitions no long range order is
building up in two particle correlation functions, since the slowing
down of the dynamics seems not to be caused by a growing correlation
length. Instead the physical origin for the slow dynamics is the
local hindering of the particle motion, the so-called cage-effect,
i.e. it is due to a mechanism operating on the {\it microscopic}
scale. However, the consequences of this local slowing down can of
course also be detected on mesoscopic and macroscopic length scales,
e.g. in thermodynamic derivatives, as soon as the time scale of the
experiment is comparable to the time scale of the structural relaxation.

In Ref.~\cite{gotze89} a formally exact method to derive frequency
dependent constitutive equations and thermodynamic derivatives in the
canonical ensemble was introduced, which thus allows to investigate
the consequences of the slowing down of the structural relaxation on
thermodynamic and hydrodynamic quantities. The main technical tools in
that paper are thermodynamic response theory and projection operator
techniques. In such an approach the most important physical ingredient is
the appropriate microscopic definition of the temperature $\theta({\bf
r}_i, {\bf p}_i)$ as a function  of  the phase space variables ${\bf
r}_i$ and ${\bf p}_i$. From a {\it mathematical} point of view the choice
of a such a function is not unique since the only condition one has to
fulfill is that
\begin{equation}
\langle \theta({\bf r}_i, {\bf p}_i) \rangle = T\quad, 
\label{eq2}
\end{equation}
where $T$ is the macroscopic temperature. For the case of supercooled
liquids also {\it physical} considerations impose restrictions to the
possible set of functions: In the canonical ensemble the temperature of
the system is imposed by an external heat bath. Therefore the definition
of the phase space function $\theta({\bf r}_i, {\bf p}_i)$ should allow
that Eq.~(\ref{eq2}) holds also in the case that $T$ is changing on a time
scale that is shorter than the one of the structural relaxation. (Such
rapid changes occur, e.g., in the earlier mentioned experiments on the
frequency dependent specific heat. But also for the case of a glass below
$T_g$ the concept of a temperature is useful, despite the fact that the
true relaxation time of the system is exceedingly large.) Since structural
relaxation is related to rearrangements in space, it can be expected that
definitions of $\theta({\bf r}_i, {\bf p}_i)$ that involve the positions
${\bf r}_i$ do not fulfill the requirement of being ``fast''. Momentum,
instead, can quickly be exchanged between particles, and hence is
transfered easily through the system. An appropriate microscopic
definition of a temperature, obeying the mentioned mathematical and
physical requirements,  can be constructed with the help of the kinetic
energy of a particle,
\begin{equation}
\theta({\bf r}_i, {\bf p}_i) \equiv \theta({\bf p}_i) = \frac{1}{3 k_B m}
{\bf p}^2_i \quad .
\end{equation}
The canonical average $\langle \theta({\bf p}_i)\rangle$ is, as required,
the macroscopic temperature $T$.  However, if one wants to define a {\it
local} density related to the temperature, it can of course not be avoided
that there is a dependence on spatial variables. Thus the best one can do
is to make this dependence as simple as possible so that it is possible
to extract the fast part easily. For a classical system a convenient
microscopic definition of a local temperature field is therefore
\begin{equation}
 T({\bf r},t) = \sum_i  \theta({\bf p}_i)
\delta({\bf r}-{\bf r}_i(t)) \quad .
\label{eq3}
\end{equation} 
The canonical average $\langle T({\bf r},t)\rangle$ is given by $n T$,
where $n = N/V$ is the number density. Although it is not a conserved
quantity, in supercooled liquids the temperature field $T({\bf r},t)$
has a slow component, due to its dependence on the positional degrees
of freedom. However, the fast fluctuation part can easily be obtained
by projecting out these slow density fluctuations:
\begin{equation}
\delta T({\bf r},t) = T({\bf r},t) 
- T \rho({\bf r},t) 
\label{eq4}
\end{equation}
or in Fourier space 
\begin{equation} 
\delta T_q(t) = T_q(t) - T  \rho_q(t)\quad .
\label{eq5}
\end{equation}
Note that by construction this temperature fluctuation does not have any
static coupling to functions of the form $G_q({\bf r}^N) = G({\bf r}^N)
\exp(i {\bf q} \cdot {\bf r}_i)$, i.e. that depend only on the positions of
the particles:
\begin{equation}
\langle \delta T_q^* G_q({\bf r}^N)\rangle =0 \quad,
\label{eq6}
\end{equation}
which shows that its dynamic autocorrelation function
$\Phi_{TT}(q,t)=(T_q(t)|T_q)$ cannot be proportional to any multi-point
density correlation function.  (For later use we have introduced the
notation $(A|B):=\langle \delta A^* \delta B \rangle$.)  Although it is
not possible to prove from the Ansatz~(\ref{eq4}) that $\delta T_q(t)$
is really a fast variable for $q>0$, i.e. that it does not undergo a
structural arrest in the canonical ensemble, this is something that can
be checked in simulations and below we will demonstrate that this is
indeed the case.  Note, however, that the {\it time} dependence of the
temperature fluctuations is of course affected by the slowing down of
structural relaxation. This was derived in \cite{gotze89} within linear
response theory. In the following we will briefly sketch the main steps in
that derivation in order to clarify the physics behind the final results.

Since in this paper we are interested in the frequency dependent
specific heat per particle at constant volume, the first thing to do is
to derive a microscopic expression for this quantity and to relate it to
the local fluctuations of the temperature. For the sake of simplicity we
will focus on the case of temperature fluctuations for vanishingly small
wave-vectors, since in this limit they decouple from the single particle
density and current fluctuations. (However, in the last part of this
section we will also consider finite wave-vectors.) To determine $c_v(z)$
we have to calculate, in analogy to the {\it static} case $\Delta E =
c_v^{eq} \Delta T$, the effect of temperature fluctuations on fluctuations
of the total energy. For this let us assume that we impose adiabatically
at time $t=0$ a (small) temperature fluctuation $\delta T_q^*(t)$. In this
case the initial probability distribution in the canonical ensemble is
\begin{equation}
\frac{ \exp (- \beta(H - h \delta T_q))}{\mbox{Tr}
\exp (- \beta(H - h \delta T_q))}\quad ,
\label{eq8}
\end{equation} 
where $h$ is the conjugate field of $\delta T_q$ and $\beta=1/k_B T$. 

In the linear response regime the time evolution of any variable
$Y^*(t)$ is given by
\begin{equation} 
\langle Y^*(t)\rangle_{NE} = \langle Y^*(t)\rangle + h \; \beta
(Y(t)|T_q) \quad,
\label{eq9}
\end{equation}
where the average on the lefthand side is done in the non-equilibrium 
ensemble (\ref{eq8}), and the one on the righthand side are performed
in the standard canonical ensemble ($h =0$ in Eq. (\ref{eq8})).
Using Eq.~(\ref{eq9}) for the case $Y = T_q$ we thus obtain for the
time dependence of the temperature fluctuation
\begin{equation}
\langle \delta T_q^*(t) \rangle = h \; \beta (T_q(t)|T_q) =
h \; \beta \Phi_{TT}(t) \quad ,
\label{eq10}
\end{equation}
and the relaxation of a fluctuation of the energy, Eq.~(\ref{eq1})
with $X_i = E_i$, is given by

\begin{equation}
\langle \delta E_q^*(t) \rangle = h \; \beta (E_q(t)|T_q)\quad .
\label{eq11}
\end{equation}

In analogy to the static case we define the frequency dependent specific
heat $c_v(z)$ as the coefficient between the fluctuation in the energy
and the one in temperature:
\begin{equation}
\langle \delta E_q^*(z) \rangle = c_v(z) \langle \delta T_q^*(z) \rangle \quad ,
\label{eq12}
\end{equation}
where $\delta E_q(z)$ and $\delta T_q(z)$ are the Laplace transforms of
$\delta E_q(t)$ and $\delta T_q(t)$, respectively. Hence we now have to
express $(E_q(t)|T_q)$ from Eq.~(\ref{eq11}) as a functional of $\delta
T_q(t)$.  The Laplace transform of $(E_q(t)|T_q)$ is $(E_q|R(z)|T_q)$,
with the resolvent $R(z) = 1/({\cal L}-z)$ and the Liouville operator
${\cal L}$.  Using a projection operator formalism with the operator
$P = |\delta T_q)(\delta T_q|\delta T_q)^{-1}(\delta T_q|$ to project on
temperature fluctuations, the correlator $(E_q|R(z)|T_q)$ can be written
as \cite{gotze89}
\begin{equation}
(E_q|R(z)|T_q) = \left\{(E_q|T_q) - (E_q|R'(z)|{\cal L} T_q)\right\}
\frac{1}{S_{TT}^c}
\langle \delta T_q^*(z) \rangle \quad .
\label{eq13}
\end{equation}
Here $R'(z):= Q/(Q{\cal L}Q-z)$, with $Q=1-P$, is the reduced
resolvent which describes the dynamics in the directions of phase
space perpendicular to temperature fluctuations (and by using the more
general formalism from Ref. \cite{gotze89} also perpendicular to density
and current fluctuations). The quantity $S^c_{TT}$ is defined as
\begin{equation}
S^c_{TT}=\langle \delta T_q^*(0) \delta T_q(0)\rangle=2NT^2/3 \quad ,
\label{eq7b}
\end{equation}
and is independent of $q$. (The superscript $c$ indicates the
canonical ensemble.) From Eq.~(\ref{eq13}) the expression for the
specific heat can be read off. We introduce the projection operators
$P_{\rho}:=|\rho_q)(\rho_q|\rho_q)(\rho_q|$ and $Q_{\rho}:=1-P_{\rho}$
and use the equations
\begin{equation}
{\cal L} Q_\rho E_q^K = - {\cal L} Q_\rho E_q^P + {\cal O}(q) = - 
{\cal L} Q E_q^P
\label{eq14}
\end{equation}
\begin{equation}
\frac{-Q{\cal L}Q}{z-Q{\cal L}Q} = Q - \frac{z Q}{z-Q{\cal L}Q} \quad,
\label{eq15}
\end{equation}
where $E_q^K$ and $E_q^P$ are the Fourier transform of the kinetic
and potential energy fields $E^K({\bf r})$ and $E^P({\bf r})$,
respectively~\cite{footnote6}. The specific heat can now be expressed as
\begin{equation}
c_v(z) = c_v^{eq} + \frac{\beta }{NT} \lim_{q\to0}
z (E_q^P|R'(z)|E_q^P)\quad ,
\label{eq16}
\end{equation}
where $c_v^{eq} = c_v=\frac{\beta }{NT} \lim_{q \to 0} (E_q|Q_\rho|E_q)$
is the equilibrium value of the specific heat \cite{schofield66} per
particle.  We emphasize, that Eq. (\ref{eq16}) is an exact expression for
the frequency dependent specific heat at constant volume. It only relies
on Eq.~(\ref{eq4}) as a physical reasonable definition of temperature
fluctuations.

Expression (\ref{eq16}) reveals the physical origin of the frequency
dependence of the specific heat very clearly: During the process of
structural relaxation the system probes the potential energy landscape.
This dynamics is governed by the (slow) relaxation of the densities and
their higher order products and hence gives rise to a nontrivial slow
dynamics even in the projected dynamics $R'(z)$, although the latter
does not contain any hydrodynamic poles.

Expression (\ref{eq16}) is the ideal starting point for approximation
schemes like the mode coupling theory. However, for exact calculations
of $c_v(z)$, e.g. in computer simulations, it cannot be used, since
it is not possible to implement the projected dynamics. To solve this
problem we will now express $c_v(z)$ in terms of correlation functions
that can be measured in a real dynamics, such as $\Phi_{TT}(q,t)$. For
this we make use of Refs.~\cite{gotze89,latz91} where similar calculation
as just presented for $c_v(z)$ were done for $c_p(z)$, the specific heat
at constant pressure, as well as the frequency dependent heat conduction
coefficient $\tilde{\lambda}(z)$. Using these results it can be shown
that for small wave-vectors the Laplace transform of $\Phi_{TT}(q,t)$
obeys the exact equation of motion
\begin{eqnarray} 
\Phi_{TT}(q,z) &:=& i \int_0^t dt \exp(i z t) 
\langle \delta T_q^*(t) \delta T_q(0)\rangle  \nonumber \\
&=&  \frac{-S^c_{TT} 3 k_B/2 }{\displaystyle z
c_p(z) + \tilde{\lambda}(z) q^2 - z^3 \frac{c_p(z) - 
c_v(z)}{z^2 - q^2 K_B(z)/m n }} \quad ,
\label{eq7}
\end{eqnarray}
where $K_B(z)$ is the frequency dependent bulk modulus and $n$
is the particle density~\cite{gotze89,latz91}. Although in general
$\tilde{\lambda}$ depends on frequency $z$, theoretical arguments
show~\cite{gotze89} that in supercooled liquids this dependence is only
weak, in agreement with experimental findings~\cite{birge86}. Therefore
$\tilde{\lambda}(z)$ can be replaced by its static value,
$\tilde{\lambda}(z =0) = i \lambda/n$, where $\lambda$ is the equilibrium
heat conduction coefficient~\cite{footnote}.  Note that the form for
the equation for $\Phi_{TT}(q,z)$ is exactly as the one in linearized
hydrodynamics \cite{forster83} except that now the transport coefficients
{\it and} the thermodynamic derivatives are generalized to functions of
the complex frequency $z$.

In the limit of vanishing wave-vectors we thus obtain from Eq.~(\ref{eq7})
the following relation between $\Phi_{TT}(q,z)$ and $c_v(z)$:
\begin{equation} 
\lim_{q \to 0} \frac{\Phi_{TT}(q,z)}{S_{TT}^c} = \frac{-3 k_B}{2 z
  c_v(z)}\quad .
\label{eq17}
\end{equation}

We emphasize that $c_v(z)$ as it occurs in Eqs.~(\ref{eq7}) and
(\ref{eq17}) is exactly the dynamic specific heat from Eq.~(\ref{eq12})
that relates fluctuations in temperatures to the ones in energy.
Using Eq.~(\ref{eq17}), we thus can express this dynamic specific
heat by a standard correlation function:

\begin{equation}
c_v(z) = \frac{-3 k_B S_{TT}^c}{2 z  \Phi_{TT}(q=0,z)} \quad .
\label{eq18}
\end{equation}

Therefore we find that in the canonical ensemble the specific heat can be
expressed by the autocorrelation function of the temperature fluctuations,
a quantity which is readily accessible in simulations. (An equivalent
definition of the specific heat in terms of potential energy fluctuations
is given in \cite{franosch00}).

It is important to note that the expression (\ref{eq7}) for $\Phi_{TT}$
is valid only in the {\it canonical} ensemble. To obtain the corresponding
relation in the {\it microcanonical} ensemble, a general transformation
due to Lebowitz {\it et al.} can be used \cite{lebowitz67}, which relates
averages of time dependent functions in the canonical ensemble to their
averages in the microcanonical ensemble.  Denoting $\delta T(q=0)$
by $\delta T_0$ we thus obtain:
\begin{eqnarray}
\Phi_{TT}(t)=\langle \delta T_0^*(t) \delta T_0 \rangle_c &=& 
\langle \delta T_0^*(t) \delta T_0 \rangle_{mc} + 
\frac{\beta^2 k_B}{N c_v^{eq}}
\left|\frac{\partial \langle T_0\rangle}{\partial \beta}\right|^2\nonumber \\ 
&=&   \Phi_{TT}^{(mc)}(t) +  \frac{N  \beta^2 }{ k_B c_v^{eq}}  
\left(\frac{\partial 1/\beta}{\partial \beta}\right)^2 \label{eq19}\\
&=& \Phi_{TT}^{(mc)}(t) +  \frac{ N }{c_v^{eq} k_B \beta^2 } \nonumber \quad .
\end{eqnarray}
Here we used, that $\langle T_0 \rangle = N T$~\cite{footnote5}. 
If we set $t=0$, the
equilibrium specific  heat $c_v^{eq}$ can thus be written as
\begin{equation}
\frac{3k_B}{2c_v^{eq}}=1-K(0) \quad ,
\label{eq20}
\end{equation}
where we have introduced the normalized temperature autocorrelation function
\begin{equation} \label{eq22}
K(t):= \Phi_{TT}^{(mc)}(t)/S_{TT}^c \quad .
\end{equation}
Putting this back into Eq.~(\ref{eq19}) and using Eq.~(\ref{eq18})
we finally obtain
\begin{equation}
c_v(z)= \frac{ 3 k_B/2}{1 -
K(t=0) - z LT[K(t)](z)}  \quad ,
\label{eq21}
\end{equation}
where $LT$ stands for the Laplace transform.

Equation (\ref{eq21}), with a slightly different definition of the
function $K(t)$, was first used in \cite{grest87}  as an ad hoc
generalization of Eq. (\ref{eq20}) to non-vanishing frequencies
and later derived within thermodynamic response theory by Nielsen
\cite{nielsen99}. 

So far we have focused on the autocorrelation function of $T_q$ in the
limit $q\to 0$. In the following we will now consider the case $q>0$. From
Eq.~(\ref{eq7}) it follows immediately that for small but not vanishing
$q$ the decay of $\Phi_{TT}(q,t)$ is dominated by the hydrodynamic pole
at $\omega = - i D_T q^2$ with $D_T = \lambda/(n c_p^{eq})$. Below we
will see that it is  important to consider also corrections of order
$q^2$ to $D_T$. As it will turn out, the frequency dependence of the
specific heat will give the main contribution in these corrections. From
theoretical considerations \cite{gotze89} as well as from experiments
\cite{birge86} we know that the heat conduction coefficients do not (or
only very weakly) depend on frequency.  In analogy to the specific heat
at constant volume, $c_v(z)$, the specific heat at constant pressure,
$c_p(z)$, will also depend on frequency. Using a generalization of
Legendre transformations between different thermodynamic ensembles it can
be shown that $c_p(z)$ can be written in the form $c_p(z) = c_p^{eq} +
z \Delta(z)$, where $\Delta(z)$ is a frequency dependent function with a
positive spectrum~\cite{latz91}. For $z= \omega + i \epsilon$, it is given
by $\lim_{\epsilon \to 0}\Delta(\omega + i \epsilon) = \Delta'(\omega)
+ i \Delta''(\omega)$.  The value of $\Delta''(\omega)$ at $\omega=0$
has several contributions, one being proportional to the longitudinal
viscosity $\eta_l = \eta_V + \frac{4}{3} \eta_S$, where $\eta_V$ and
$\eta_S$ are the bulk and shear viscosity, respectively. The relaxation
time $\tau_q$ of $\Phi_{TT}(q,t)$ which is related to the hydrodynamic
pole can be calculated by determining the frequency at which the main
denominator of Eq. (\ref{eq7}) vanishes up to order $q^6$. Making the
Ansatz $1/\tau_q=\omega_R = - i D_T q^2 + i B q^4 - A q^4$   we obtain
\begin{equation} \label{rayleigh}
\omega_R = - i D_T q^2 \, \left ( 1  -
  \frac{\Delta''(\omega=0)}{c_p^{eq}} D_T q^2 +  \frac{\gamma -1}{K_B
    \gamma/mn}  D_T^2 q^2 \right )   + D_T^2 \frac{\Delta'(\omega
  =0)}{c_p^{eq}}  q^4 \quad ,
\end{equation}
where $\gamma$ is $c_p^{eq}/c_v^{eq}$,  $\gamma -1$ is the Landau -
Placzek ratio, and $K_B$ is the static bulk modulus.  The real part indicates an
experimentally undetectable shift of the position of the Rayleigh peak to
finite frequencies. Since it is the frequency of an overdamped harmonic
oscillation,  it does not influence the relaxation time $\tau_q$ of
the correlator in real time.  From Eq. (\ref{rayleigh}) the relaxation
time up to order $q^0$ of the exponentially decaying temperature 
autocorrelation function is thus given by

\begin{equation}
\tau_q= \frac{1}{\omega_R} = \frac{n c^{eq}_p}{\lambda} \frac{1}{q^2} +
\frac{\Delta''(\omega=0)}{c_p^{eq}} - D_T \frac{\gamma -1}{K_B\gamma/mn}
\; + {\cal O}(q^2)\quad .
\label{eq23}
\end{equation}

It is interesting to note that for the case of supercooled liquids the
term proportional to $q^0$ is only positive due to the existence of a
frequency dependent specific heat, i.e. that $\Delta''>0$. Without it,
the subtraction of the positive Landau-Placzek ratio $\gamma -1$ would
lead to a negative offset in a plot of $\tau_q$ against $1/q^2$.  Below we
will check the validity of this $q-$dependence and calculate from this
relation the value of $\lambda$. Note that in the case of {\it finite}
wave-vectors the autocorrelation functions $\langle \delta T_q^*(t)
\delta T_q(0) \rangle$ in the canonical and microcanonical ensemble
are identical, and hence it is not necessary to use the transformation
formula of Lebowitz {\it et al}~\cite{lebowitz67}.

\section{Model and Details of the Simulations}
\label{sec3}

In this section we discuss the system we used to test the ideas presented
in the previous section and give the details of the simulations. At the
end we also briefly discuss the influence of finite size effects on
the results.

Although the formalism presented in the previous section is of course
applicable to all types of glass forming liquids, it is of special
interest to investigate a system which exists also in reality, since
this opens the possibility to compare the results from the simulations
with those from experiments. To do a simulation of a real material
one needs a potential that describes reliably the interactions between the
atoms of this substance. In the case of silica such a potential does
indeed exist, since a decade ago van Beest {\it et al.} (BKS) used
{\it ab initio} calculations to obtain a classical force field for this
material~\cite{beest90}. The functional form of this potential is given by
\begin{equation}
\phi_{\alpha \beta}(r)=
\frac{q_{\alpha} q_{\beta} e^2}{r} + 
A_{\alpha \beta} \exp\left(-B_{\alpha \beta}r\right) -
\frac{C_{\alpha \beta}}{r^6}\quad \alpha, \beta \in
[{\rm Si}, {\rm O}],
\label{eq50}
\end{equation}
where $r$ is the distance between two ions of type $\alpha$
and $\beta$.  The value of the constants $q_{\alpha}, q_{\beta},
A_{\alpha \beta}, B_{\alpha \beta}$, and $C_{\alpha \beta}$ can be
found in Ref.~\cite{beest90}. The short range part of the potential was
truncated and shifted at a distance of 5.5\AA, which leads to a better
agreement of the results for the density of vitreous silica as predicted
from this model with the experimental values~\cite{vollmayr96a}. In
the past it has been shown that this potential is able to reproduce
reliably various properties of amorphous silica, such as its structure,
its vibrational and relaxational dynamics, the static specific
heat below the glass transition temperature, and the conduction of
heat~\cite{vollmayr96a,vollmayr96b,koslowski97,taraskin97a,taraskin97b,horbach98a,horbach98b,horbach98c,jund99,horbach99,ispas99,horbach_boson}.
Thus it is reasonable to assume that this potential will also reproduce
well the quantities needed to calculate the frequency dependent specific
heat, i.e. the correlation function $K(t)$ in Eq.~(\ref{eq22}).

Using the BKS potential, the equations of motion were integrated with
the velocity form of the Verlet algorithm with a time step of 1.6fs. The
sample was first equilibrated by coupling it every 50 time steps to a
stochastic heat bath for a time which allowed the system to relax at
the temperature of interest. (We assumed a sample to be relaxed if the
coherent intermediate scattering function at the wave-vector corresponding
to the main peak in the static structure factor had decayed to zero within
the time span of the equilibration~\cite{hansen_mcdonald86}.) After this
equilibration we started a microcanonical run for the production from
which we determined the equilibrium dynamics at various temperatures. The
temperatures investigated were 6100K, 4700K, 4000K, 3580K, 3250K,
3000K, and 2750K. At the lowest temperature $K(t)$ decays so slowly that
runs with 30 million time steps were needed to equilibrate the sample,
which corresponds to a real time of 49ns. All the simulations were done
at a constant density of 2.36g/cm$^3$ which corresponds to a pressure
around 0.87~GPa\cite{scheidler_dipl}. Since the temperature expansion
coefficient of silica is small~\cite{vollmayr96a,mazurin83}, it can be
expected that a constant pressure simulation would basically give the
same results as our simulation.

To obtain also results in the glass at room temperature we cooled the
sample from 3000K to 2750K with a cooling rate around $3\cdot 10^{11}$K/s,
and then rapidly quenched it to 300K. The so obtained glass was annealed
for additional 300,000 time steps before we started the measurement of
the various quantities.

Since the main observable of interest, $K(t)$, is a collective variable,
it has a relatively large statistical error. Therefore we averaged all
the runs over 200 independent samples for $T\geq 4700$K, over 100 samples for
$4000$K $\geq T \geq 3000$K, and 50 samples for $T=2750$K. In order to make so
many independent runs the system size had to be rather small. Therefore
we choose a system of 112 silicon and 224 oxygen ions in a cubic box
of size 18.8\AA. The Coulombic part of the potential was evaluated
with the Ewald method and a parameter $\alpha$ of 7.5\AA$^{-1}$. All
these calculations took around 8 years of single CPU time on a parallel
computer with high end workstation processors.

As it was shown in Refs.~\cite{horbach_boson,horbach96}, the {\it dynamics}
of network glass formers shows quite strong finite size effects since the
small systems lack the acoustic modes at small wave-length. Thus it can
be expected that if we determine $K(t)$ for a system of 336 particles
the result will be different from the ones for a system of macroscopic
size. In order to check the influence of the system size on our results
we have made some test runs with 1008 particles and found, in agreement
with the results of Refs.~\cite{horbach_boson,horbach96}, that at long
times the dynamics of the larger system is a bit faster than the one of
the small one~\cite{scheidler_dipl}. However, the qualitative behavior
of the various relaxation functions are independent of the system size
and therefore we can expect that the results presented in this work will
hold also for larger systems.

\section{Results}
\label{sec4}

In this section we present the results from our simulation, i.e. the
temperature dependence of $K(t)$ and the frequency dependence of the
specific heat. At the end we briefly discuss the time and temperature
dependence of the generalization of $K(t)$ to finite wave-vectors.

\begin{figure}[h]
\psfig{figure=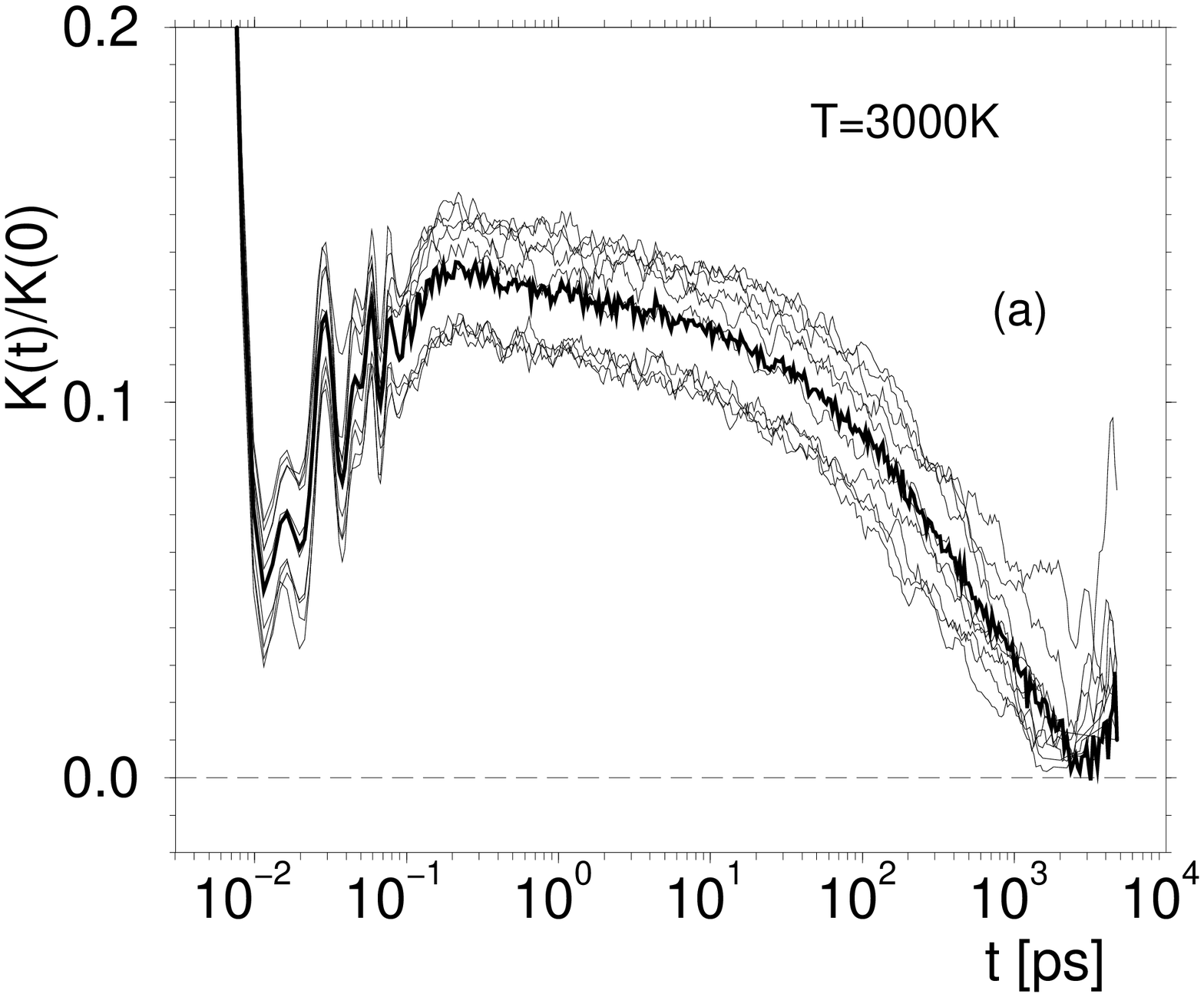,width=8cm,height=6.5cm}
\vspace*{1mm}

\psfig{figure=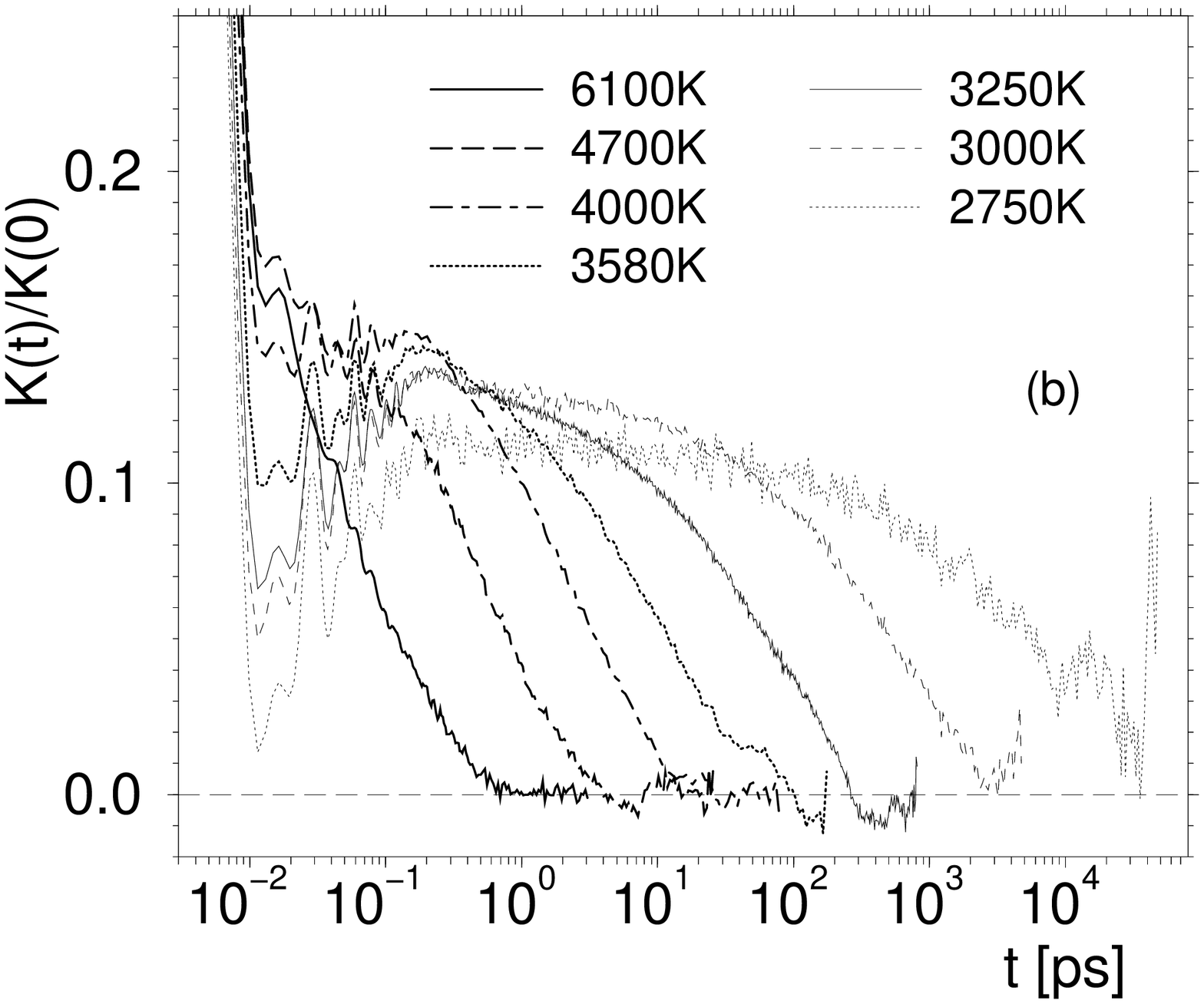,width=8cm,height=6.5cm}
\psfig{figure=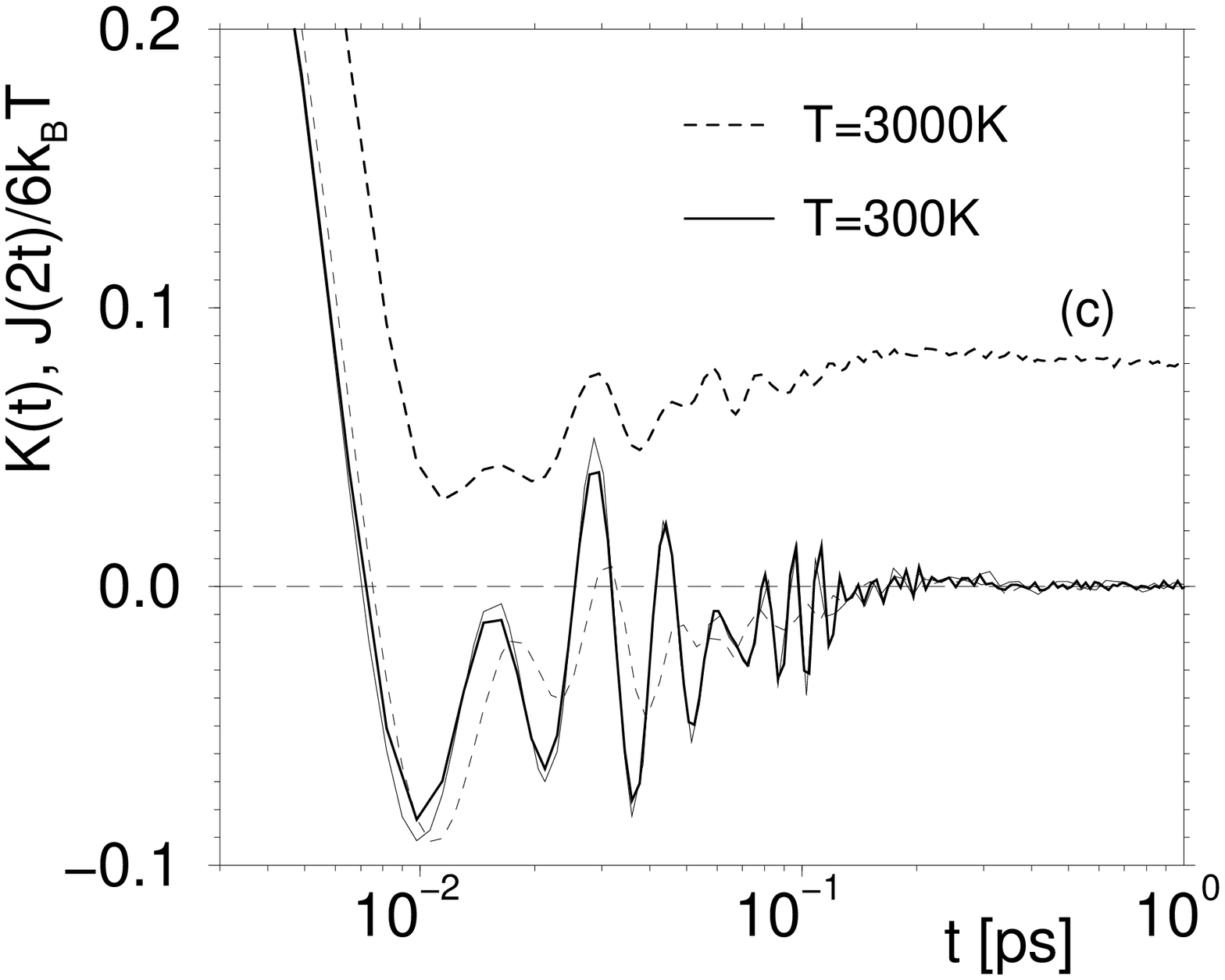,width=8cm,height=6.5cm}
\vspace*{2mm}

\caption{Time dependence of the normalized kinetic energy autocorrelation
function. a) $T=3000$K; each thin curve is the average over ten
different samples; the average over these curves gives the bold curve. b)
$K(t)/K(0)$ for all temperatures investigated. c) Bold curves: $K(t)$
for $T=3000$K and $T=300$K; thin curves: velocity-autocorrelation function
$J(2t)/6k_BT$ (see Appendix A) at the same temperatures.
} 
\label{fig1}
\end{figure}

As it is obvious from Eq.~(\ref{eq21}), the first step in the calculation
of the frequency dependent specific heat is to determine $K(t)$, the
autocorrelation function of the fluctuation in the kinetic temperature.
Since $K(t)$ is a collective quantity it is relatively difficult of
determine it with high accuracy. This is shown in Fig.~\ref{fig1}a where
we plot $K(t)/K(0)$ at $T=3000$K. Each of the thin curves corresponds to
the average over ten independent samples. From the figure we recognize
that even if the shape of the different curves is quite similar, their
height varies considerably. Thus one should realize that even if the
average over 100 samples gives a quite nice and smooth curve (bold
solid curve) it still might be subject to a significant statistical error.

In Fig.~\ref{fig1}b we show the time dependence of $K(t)/K(0)$
for all temperatures investigated. From this figure we see that
for all temperatures this correlator decays within $10^{-2}$ps
to a value smaller than 0.2, i.e. very rapidly. For the highest
temperatures $K(t)$ then shows a small shoulder at around 0.01ps, a
feature which, at these temperatures, is not observed in correlation
functions such as the intermediate scattering function $F(q,t)$,
where $q$ is the wave-vector~\cite{horbach98b,horbach99}. (See
Ref.~\cite{hansen_mcdonald86} for a definition of $F(q,t)$). With
decreasing temperature this shoulder extends to larger and larger times
until we observe at the lowest temperature a well defined plateau which
extends over several decades in time. Thus from this point of view $K(t)$
behaves qualitatively similar to $F(q,t)$. Since for this quantity it is
customary to refer to this final decay as the ``$\alpha-$relaxation''
we will use the same term in the case of $K(t)$ as well. Note that the
height of the plateau in $K(t)$ is only around 0.1, which shows that
most of the correlation of the kinetic energy is lost during the brief
ballistic flight of the particles, $t \leq 10^{-2}$ps, inside the cage.

From the figure we also see that with decreasing temperature the
correlation function starts to show a minimum at short times. The reason
for this feature is that at low temperatures and short times the system
behaves similar to a harmonic solid and thus it can be expected that
the connection between $K(t)$ and the velocity-autocorrelation function
$J(t)$, see Eq.~(\ref{eqa4}) in the Appendix, starts to hold. (We remind
the reader that in supercooled liquids the velocity-autocorrelation
function shows a dip at short times.) That this is indeed the case is
demonstrated in Fig.~\ref{fig1}c, where we compare the two correlators
at a high and a low temperature. We see that for $T=300$K, i.e. in the
glass where the harmonic approximation is valid, $K(t)$ and $J(2t)$ are
identical within the accuracy of the data (solid lines). For temperatures
at which the system is still able to relax the situation is different. At
short times $K(t)$ (bold dashed line) shows oscillations with extrema
which are located at the same times at which also $J(2t)$ (thin dashed
line) shows maxima and minima. Thus the harmonic-like character of the
motion on these time scales is clearly seen. For larger times, however,
$J(2t)$ goes rapidly to zero whereas $K(t)$ shows the above discussed
plateau before it decays to zero at very long times.

\begin{figure}[h]
\psfig{figure=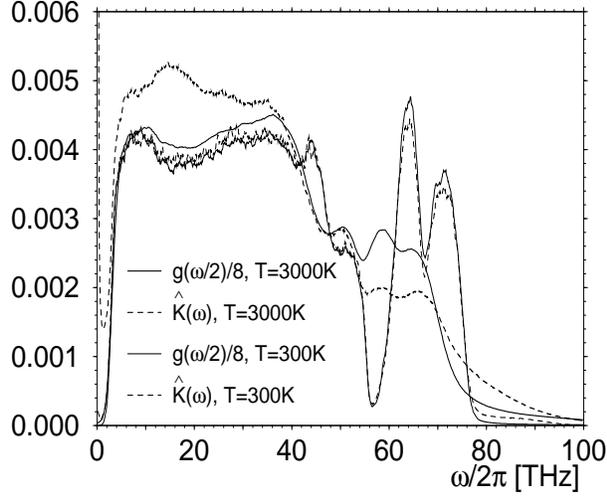,width=8cm,height=6.5cm}
\vspace*{2mm}

\caption{Frequency dependence of $\hat{K}(\omega)$ and $g(\omega/2)/8$ 
(dashed and solid curves, respectively). Bold lines: $T=3000$K, 
thin lines: $T=300$K.
}
\label{fig2}
\end{figure}

In order to investigate this point in more detail we can make use of
Eq.~(\ref{eqa5}), which relates the spectrum of $K(t)$, $\hat{K}(\omega)$,
to $g(\omega)$, the time Fourier transform of the velocity-autocorrelation
function. At low temperatures the latter quantity is nothing else than
the density of states of the system. (In order to calculate these
Fourier transforms we made use of the Wiener-Khinchin theorem which
relates the power spectrum of a time dependent signal to the Fourier
transform of the corresponding autocorrelation function~\cite{press84}.)
In Fig.~\ref{fig2} we show $\hat{K}(\omega)$ for a temperature in
the melt and in the glass (dashed lines) and compare these curves
with $g(\omega/2)/8$ at the same temperatures (solid lines). We see
that, within the accuracy of our data, in the glass the two curves
are indeed identical. (We remind the reader that the two peaks at high
frequencies correspond to intra-tetrahedral motion of the atoms, whereas
the broad band at lower frequencies corresponds to (mostly) delocalized
inter-tetrahedral motion~\cite{galeener79,pasquarello98}.) For the case of
the melt, however, the two curves differ significantly from each other.
Although the general shape of $\hat{K}(\omega)$ and $g(\omega/2)$
are similar, the former quantity has a much smaller intensity at the
two peaks at high frequencies but a higher one in the broad band. Note
that this ``excess'' in low frequency modes has nothing to do with the
fact that $K(t)$ shows at long times a $\alpha-$relaxation, whereas
$J(2t)$ does not, since at $T=3000$K the typical frequencies of the
$\alpha-$relaxation are smaller than 1THz, and their contribution
to the spectra can only be seen as the narrow peak at $\omega=0$ (see
Fig.~\ref{fig2})~\cite{footnote4}. Thus from this figure we can conclude
that at low temperatures the harmonic approximation is very good whereas
at intermediate temperatures significant deviations are observed.

Since $c(z)$ is related to the Laplace transform of $K(t)$,
see Eq.~(\ref{eq21}), one has to calculate this transform with high
accuracy. From Fig.~\ref{fig1}b we see however, that at low temperatures
$K(t)$ extends over many decades in time which makes the calculation
of this transform a non-trivial matter.  From this figure one also
recognizes that, despite the large number of samples we used, the curves
have still a significant amount of noise, most noticeable at long times and
the lowest temperatures, since for these we used fewer samples.  Since within
the accuracy of the data the shape of the curves does, {\it in the late
$\alpha-$relaxation regime}, not depend on temperature, we substituted
for $T\leq 4700$K that part of the curve which was below 0.02 by the
corresponding part of the curve for $T=6100$K, after having it shifted to
such large times that the resulting curve was smooth at 0.02. Subsequently
the so obtained curves were smoothed and the Laplace transform calculated
making use of the formula by Filon~\cite{abramowitz64}.

\begin{figure}[h]
\psfig{figure=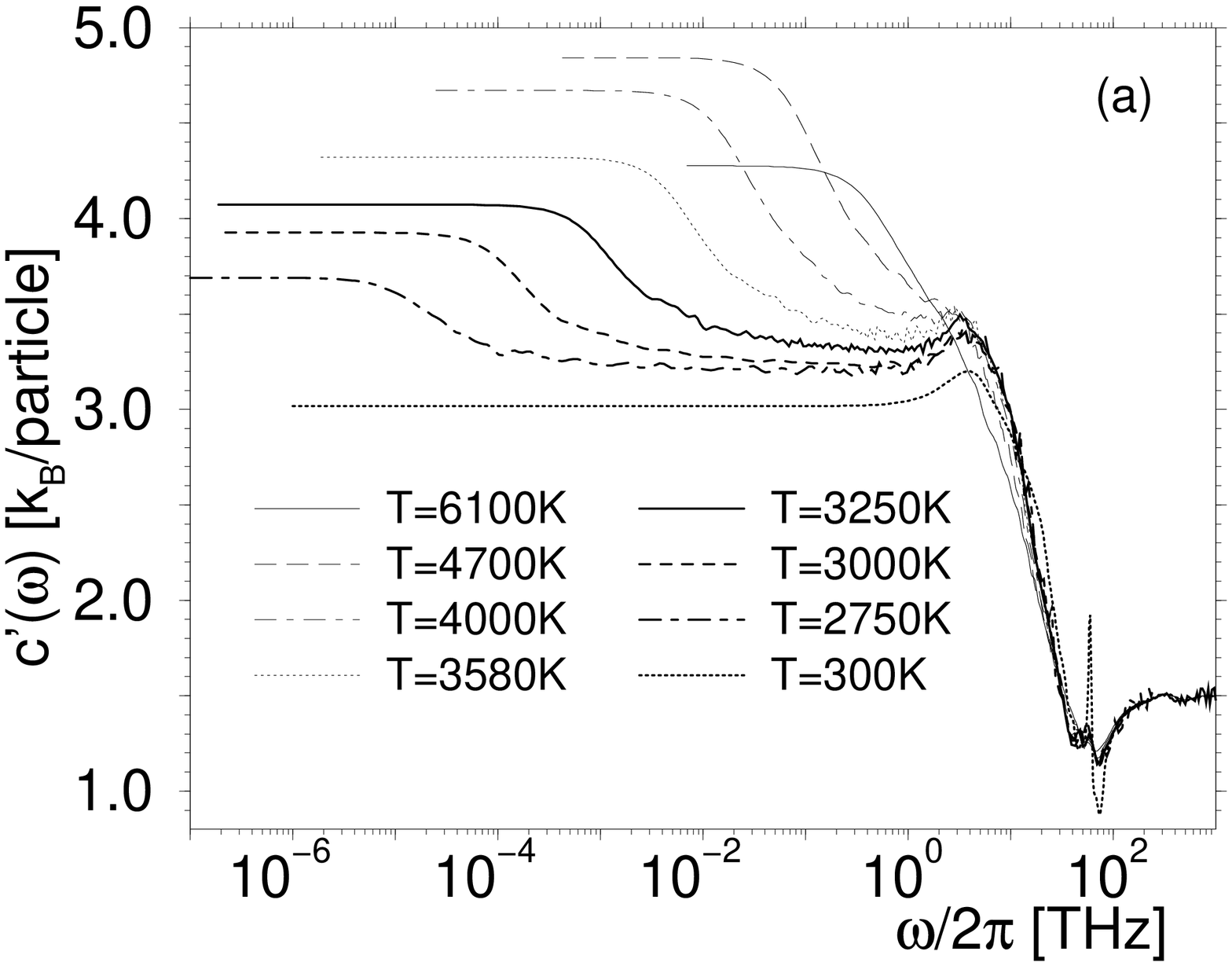,width=8cm,height=6.5cm}
\psfig{figure=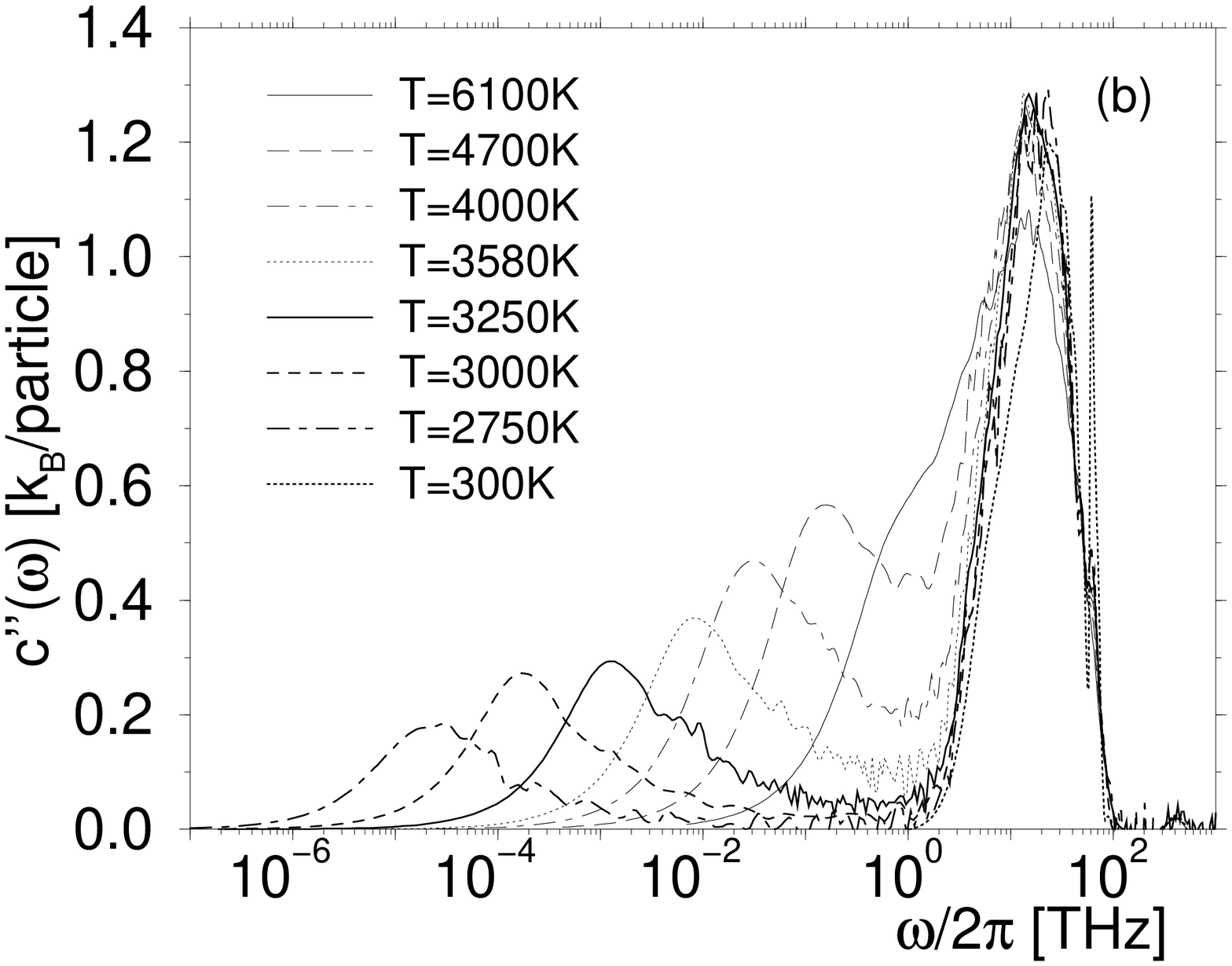,width=8cm,height=6.5cm}
\vspace*{2mm}

\caption{Frequency dependence of the specific heat for all temperatures
investigated. a) Real part. b) Imaginary part.
}
\label{fig3}
\end{figure}

In Fig.~\ref{fig3} we show the real and imaginary part of $c(z)$ for all
temperatures investigated. (In the following we will assume $z=\omega+i
\epsilon$, with $\epsilon \to 0$.) To discuss the frequency dependence
of these curves let us focus for the moment on $c'(\omega)$ for $T=2750$K,
the lowest temperature at which we could equilibrate our system. At
very high frequencies $c'(\omega)$ becomes independent of $\omega$ and has a
value of 1.5, which is the specific heat of an ideal gas. This result
is reasonable since these high frequencies correspond to times at which
the dynamics of each particle is not affected by the other ones, i.e.,
they move just ballistically. Upon lowering the frequency the specific
heat rises rapidly since we are now in the frequency regime in which
the dynamics of the system is mainly dominated by vibrations, see
Fig.~\ref{fig2}. Therefore at these frequencies the system can take
up energy giving rise to the increase of $c'(\omega)$. Note that before this
increase occurs, $c'(\omega)$ shows a little dip around 40THz, i.e., it falls
below the ideal gas value. This dip can easily be understood by the
harmonic picture proposed above, because the specific heat for a single
harmonic oscillator shows a singularity at its resonance frequency.
Since in our system we have many different oscillators that have typical
frequencies between 1 and 80THz, this singularity is smeared out and
results in the dip and subsequent strong increase of $c'(\omega)$. The same
mechanism is the reason for the little peak at around 5THz.

If the frequency is decreased further, $c'(\omega)$ stays constant until 
$\omega^{-1}$ is on the order of the time scale of the $\alpha-$relaxation,
i.e. the time scale of the structural relaxation. In this frequency
range the system is again able to take up energy and hence $c'(\omega)$
increases again. At even lower frequencies $c'(\omega)$ becomes constant,
i.e., it has reached the value of the {\it static} specific heat. This
sequence of features in $c'(\omega)$ can of course also be found in
$c''(\omega)$, since the two functions are related by the Kramers-Kronig
relation. In Fig.~\ref{fig3}b we see that at high frequencies the
imaginary part has a peak which corresponds to the vibrational excitations
in the system. At much lower frequencies we find the $\alpha-$peak
which corresponds to the structural relaxation process, i.e. the type of
dynamics in which particles change their neighbors. For future reference
we will introduce the terms ``vibrational and configurational part of the
specific heat'' by which we mean the height of the plateau in $c'(\omega)$
at intermediate frequencies and the height of the step at low frequencies,
respectively, and will denote them by $c_{\rm conf}$ and $c_{\rm vib}$.

Let us now discuss the temperature dependence of $c'(\omega)$ and
$c''(\omega)$. From Fig.~\ref{fig3}a we see that the specific heat at
intermediate frequencies is essentially independent of $T$, since the
vibrational motion of the ions is just a weak function of temperature. The
main effect of an increase in temperature is that the height of the flat
region at very small $\omega$, i.e. the static specific heat, increases,
and that the crossover from this region to the plateau at intermediate
frequencies moves to higher frequencies. At the highest temperature
this crossover frequency has moved up to such high frequencies that no
intermediate plateau is visible anymore, which means that in the system
there is no longer a separation of time scales for the vibrational
and relaxational processes.  Furthermore we find that at the highest
temperature the height of the plateau at small frequencies has decreased,
i.e. that the value of the static specific heat has decreased. This effect
is most likely related to the fact that silica shows a density anomaly,
which for the present model occurs at around 4600K~\cite{vollmayr96a}.

The temperature dependence just discussed is also found in the imaginary
part of $c(z)$ in that, with increasing temperature, the $\alpha-$peak
moves continously to higher frequencies until it merges with the
microscopic peak. All this behavior is qualitatively similar to the
one found for dynamic observables that measure the structural relaxation,
such as the dynamic susceptibility~\cite{horbach_diss,horbach00}.
Thus this gives evidence that the observable related to the structural
relaxation and $c(z)$ are closely connected to each other.

We also note that at low temperatures the form of the curves at low
frequencies as well as their temperature dependence resembles very much
the ones found in
experiments~\cite{birge85,birge86,dixon88,dixon90,menon96,jeong95}. The
main difference is that in the simulation it is possible to measure
$c(z)$ even at such high frequencies that the effect of the microscopic
vibrations becomes visible. Thus it is possible to follow continuously the
evolution of $c(z)$ from the microscopic regime to the mesoscopic one,
i.e. to investigate the whole frequency dependence of $c(z)$ from the
liquid state to the viscous state.  In contrast to this, experiments
can probe only the frequency range below $10^4$Hz and therefore only
the $\alpha-$regime is observable. However, since experimentally it is much
simpler to equilibrate the material also at temperatures close to the glass
transition temperature, $c(z)$ can be measured at significantly lower
temperatures than in a simulation.

Finally we mention that we have included in Fig.~\ref{fig3} also the
data for $T=300$K, i.e. a temperature at which the system is deep in the
glass state. We see that this curve follows the pattern of the equilibrium
curves very well in that it shows also the ``harmonic resonance'' at high
frequencies and then a plateau at lower frequencies. No second plateau
is seen in $c'(\omega)$ (or an $\alpha-$peak in $c''(\omega)$) at very
small frequencies since at this temperature these features would occur
at such low $\omega$ that they are not visible within the time span of
the simulation (or even an experiment).

\begin{figure}[h]
\psfig{figure=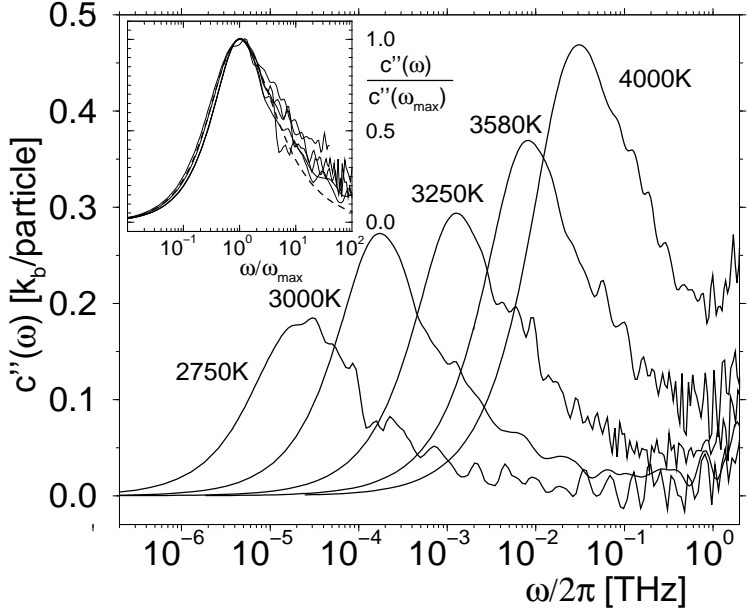,width=10cm,height=8cm}
\vspace*{2mm}

\caption{Main figure: Frequency dependence of the $\alpha-$peak for 
intermediate and low temperatures. Inset: Same curves scaled by the 
height of their maximum versus $\omega/\omega_{\rm max}$, where $\omega_{\rm
max}$ is the location of the maximum. Dashed line: 
Kohlrausch-Williams-Watts function.
}
\label{fig4}
\end{figure}

Since the part of the specific heat that is related to the structural
relaxation is the $\alpha-$relaxation peak at low frequencies, we will
study this peak now a bit in more detail. In Fig.~\ref{fig4} we show
an enlargement of this peak for intermediate and low temperatures. We
clearly see that with decreasing temperature the area under the peak
becomes smaller, which means that the configurational part of the
specific heat decreases. This temperature dependence might be somewhat
unexpected since for other dynamic susceptibilities, such as the one
connected to the intermediate scattering function, one finds that the
so-called time temperature superposition principle is valid, i.e. a
decrease in temperature just gives rise to a horizontal shift of the
$\alpha-$peak~\cite{horbach98b,horbach00}. However, since the area under
the $\alpha-$peak is related via the Kramers-Kronig relation to the height
of the step at low frequency in $c'(\omega)$, i.e. the configurational
part of the specific heat, it is not surprising that this area depends
on temperature, and below we will show that this is indeed the case.
This tendency can also be easily understood from Eq.~(\ref{eq21}): For
this we assume that in the $\alpha-$relaxation regime at low temperatures
the function $K(t)$ can be written as $K(t,T)=f(T)\tilde{K}(t/\tau(T))$,
where $f(T)$ is the height of the plateau and $\tau(T)$ is the typical
relaxation time. That this assumption is reasonable can be inferred
from Fig.~\ref{fig1}b. It is now simple to show that $c''(\omega)$
can be approximated by $c''(\omega) \approx\tilde{c}''(\omega
\tau(T))f(T)/(1-K(t=0))^2=\tilde{c}''(\omega\tau(T))
f(T)4(c_v^{eq})^2/9k_B^2$, with a master function $\tilde{c}''$. (Here
we also made use of Eq.~(\ref{eq20})).  Since the function
$\tilde{c}''$ is assumed to be independent of temperature, we see
that the whole $T-$dependence of $c''(\omega)$ is given by a shift
in frequency proportional to $\tau(T)$ and a vertical rescaling by
$f(T)(c_v^{eq})^2$. Thus we conclude that the configurational part of
the specific heat is proportional to $f(T)(c_v^{eq})^2$. For the present
system both, $f(T)$ and $c_v^{eq}$, decrease with decreasing temperature,
and thus it is clear that the same is true for $c_{\rm conf}$. However, in
certain materials, such as fragile glass-formers, it is sometimes observed
that the specific heat {\it increases} with decreasing temperature. Thus
in these cases the temperature dependence of $c_{\rm conf}$ does not have
to decrease monotonically, but it might, e.g., exhibit a local maximum.

Although the height of the $\alpha-$peak changes, its shape seems
to be independent of temperature. To demonstrate this we plot in
the inset of Fig.~\ref{fig4} $c''(\omega)/c''(\omega_{\rm max})$ versus
$\omega/\omega_{\rm max}$, where $\omega_{\rm max}(T)$ is the location of the
$\alpha-$peak. Since the curves for the different temperatures fall
on top of each other, to within the accuracy of the data, we conclude
that the shape does indeed not change with temperature. Also included in
the figure is the Fourier transform of a Kohlrausch-Williams-Watts-law
with a stretching exponent 0.9. We see that this functional form
fits the master curve quite well, at least if one does not go to too
high frequencies. At these higher frequencies the scaling breaks down
due to the presence of the microscopic peak. We also mention that the
(relatively large) value of the stretching parameter is reasonable, since
in strong glass formers the stretching in the structural relaxation is
usually weak, and indeed we have found that also for the present model
the structural relaxation shows only a weak stretching~\cite{horbach98b}.

\begin{figure}[h]
\psfig{figure=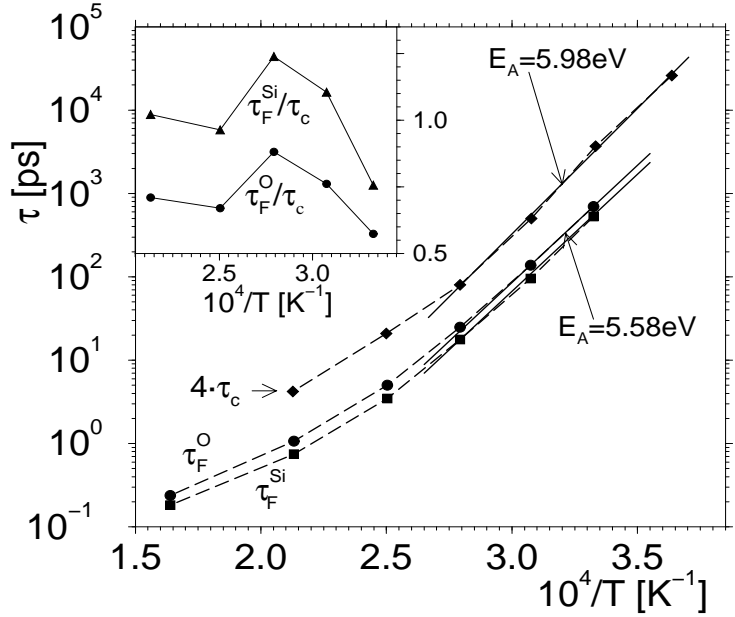,width=10cm,height=8cm} 
\vspace*{2mm}

\caption{Main figure: Temperature dependence of the $\alpha-$relaxation 
times as determined from the incoherent intermediate
scattering function and the frequency dependent specific heat. The straight
lines are fits with an Arrhenius law. Inset: ratio of these relaxation times.
}
\label{fig5}
\end{figure}

Further evidence that the structural relaxation and the frequency
dependent specific heat are closely connected to each other can be
obtained by comparing the typical time scales for these functions. For
this we have calculated the (incoherent) intermediate scattering function
$F_s(q,t)$~\cite{hansen_mcdonald86} for a wave-vector $q=1.7$\AA$^{-1}$,
which corresponds to the location of the first peak in the static
structure factor~\cite{horbach99}. We have defined the $\alpha-$relaxation
time $\tau_F^{\alpha}$, $\alpha \in \{\rm Si, O\}$, by the time it takes
this correlation function to decay to $1/e$ of its initial value. To
characterize the time scale for the specific heat we have determined from
$c''(\omega)$ the position of the maximum of the $\alpha-$peak and defined
the relaxation time $\tau_c$ as $1/\omega_{max}$. In Fig.~\ref{fig5}
we show the temperature dependence of these relaxation times in an
Arrhenius plot.  From that graph we see that the relaxation times
$\tau_c$ and $\tau_F^{\alpha}$ track each other very well in that at
low temperatures both of them show an Arrhenius law with a very similar
activation energy (numbers are given in the figure).  With increasing
temperature, deviations from this law are seen, the origin of which have
been discussed elsewhere~\cite{horbach99}, but also these deviations
are the same for both quantities. That the structural relaxation and
the specific heat do indeed track each other is demonstrated in the
inset, where we plot the ratios $\tau_F^{\alpha}/\tau_c$ as a function
of inverse temperature. Since we see that these ratios are independent
of temperature to within the accuracy of the data, we can conclude that
the temperature dependence of the three quantities is indeed the same.

\begin{figure}[h]
\psfig{figure=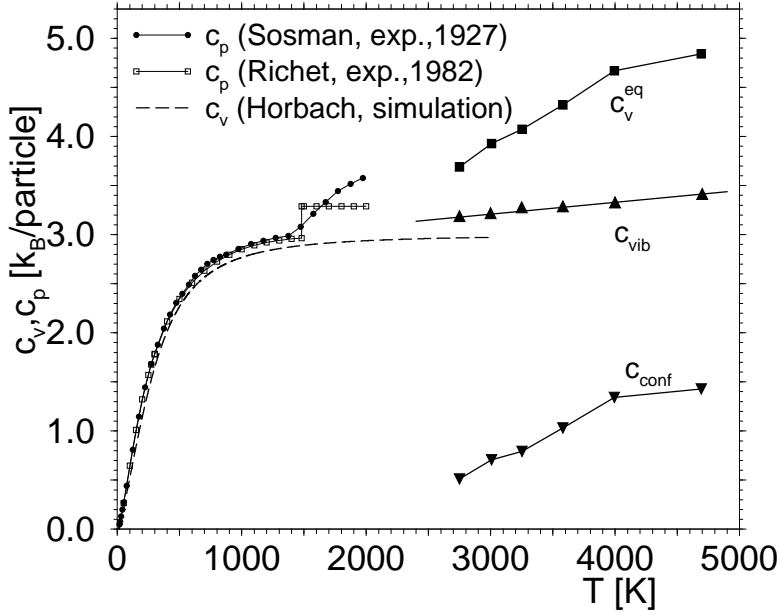,width=10cm,height=8cm} 
\vspace*{2mm}

\caption{Temperature dependence of $c_{\rm conf}$ and $c_{\rm vib}$, the
configurational and vibrational part of the specific heat, respectively,
and of their sum $c_v^{eq}$. The dashed line is the specific
heat as calculated from the harmonic
approximation~\protect\cite{horbach98c}. The curves with the small
symbols are experimental data from Ref.~\protect\cite{sosman27,richet82}.
}
\label{fig6}
\end{figure}

In the discussion of Fig.~\ref{fig3}a we have mentioned that the
increase in $c'(\omega)$ at high frequencies is due to the vibrational
degrees of freedom, whereas the step at lower frequencies is due to the
relaxation of the configurational degrees of freedom. By measuring the
heights of these two steps it thus becomes possible to determine the
contribution of the vibrational and configurational degrees of freedom
to the static specific heat and to investigate how these quantities
depend on temperature~\cite{footnote2}. The results obtained are shown
in Fig.~\ref{fig6}, where we plot $c_v^{eq}$, the value of $c'(\omega)$
at very low frequencies which is hence the static specific heat,
$c_{\rm conf}$, the height of the step in $c'(\omega)$ at low
frequencies, and $c_{\rm vib}$, the value of $c'(\omega)$ at intermediate
frequencies. Several observations can be made: Firstly, $c_{\rm vib}$
shows a very regular temperature dependence which can be approximated
well by a linear function of temperature, at least in the temperature
range investigated. An extrapolation of this temperature dependence to
lower temperatures shows that $c_{\rm vib}$ attains the harmonic value of
3$k_B$ only at low temperatures ($\approx$~1000K), i.e. at temperatures
that are well below the (experimental) glass transition temperature,
which is at 1450K~\cite{bruckner70}, a value that seems to be reproduced
reasonably well by the present model~\cite{horbach99}.  Hence we find
that $c_{\rm vib}$ is affected by anharmonic effects even at relatively
low temperatures. We also mention that the vibrational part $c_{\rm vib}$
in the temperature range 2750K~$\leq T \leq$~3500K seems to be in nice
agreement with a {\it linear} extrapolation of the experimental specific
heat of the glass below $T_g=1450$K to higher temperatures, i.e.  if one
leaves out the increase of the specific heat due to the glass transition.

The temperature dependence of $c_{\rm conf}$ is much more pronounced in than
the one of $c_{\rm vib}$, in that it shows around 4000K a crossover from
a relatively weak temperature dependence at high $T$ to a stronger one at
low $T$. Furthermore we see that $c_{\rm conf}$ is significantly smaller
than $c_{\rm vib}$, which is in agreement with the experimental result that
strong glass formers show only a small drop in the specific heat when
the temperature is lowered below the glass transition temperature, i.e.
when the relaxational degrees of freedom are frozen~\cite{angell94}. We
also note that for a strong glass former one expects that the Kauzmann
temperature $T_K$ is very small~\cite{angell85}. From the figure it
seems however, that a naive extrapolation of $c_{\rm conf}$ to lower
temperatures leads to a intercept with the temperature axis around
$T_{\rm conf}\approx 2000$K. Since the inequality $T_{\rm conf} \leq
T_K$ must hold, this type of extrapolation thus leads to an estimate of
$T_K\geq 2000$K. This high estimate of $T_K$ is corroborated by recent
results of the same model in which $T_K$ was estimated by the direct
calculation of the entropy and a subsequent extrapolation to lower
temperatures~\cite{saika00}.

These results for $T_K$ depend of course crucially on the way $c_{\rm
conf}$ is extrapolated to lower temperatures. From Fig.~\ref{fig6} it
is clear that it is also possible to make this extrapolation in such
a way that, e.g., at $T=1450$K its value is around 0.5$k_B$/particle,
i.e. equal to the height of the step in the experimental curve at
$T_g$ (see experimental curves in figure)~\cite{footnote3}. If the
extrapolation is done in this way, the estimate of $T_{\rm conf}$
is moved to much lower temperatures. Thus it will be very interesting
to attempt to do simulations at even lower temperatures in order to
minimize the effects of this extrapolation. For this it will of course
be necessary to equilibrate the system at even lower temperatures, which
is computationally difficult. One promising way to achieve this is the
so-called method of ``parallel tempering'', and work in this direction
is presently done~\cite{stuehn00,yamamoto00,kob00}.

Also included in the figure is the specific heat of the system as
calculated from the harmonic approximation~\cite{horbach98c} (dashed
line). This was done by determining the eigenvalues of the dynamic matrix,
and hence the density of states $g(\omega)$, and using the expression
\begin{equation}
c_v=\frac{h^2}{k_BT^2}\int_0^\infty \frac{\omega^2 g(\omega)
\exp(h\omega/k_BT)}{\left( \exp(h\omega/k_BT)-1\right)^2} d\omega.
\end{equation}
More details on this calculation can be found in Ref.~\cite{horbach98c}
where it has been shown that this theoretical curve agrees very well
with experimental data below the glass transition temperature $T_g$
(see the experimental data of Sosman~\cite{sosman27} and 
Richet {\it et al.}~\cite{richet82},
and the theoretical curve of Horbach {\it et
al.}~\cite{horbach98c} in the figure).  From the graph we see that an
extrapolation of $c_v^{eq}$ to lower temperatures extrapolates nicely
to the experimental data and that an extrapolation of $c_{\rm vib}$
to lower temperatures can be joined smoothly to the curve from the
harmonic approximation, thus showing that the two types of calculations
are consistent with each other.

\begin{figure}[h]
\psfig{figure=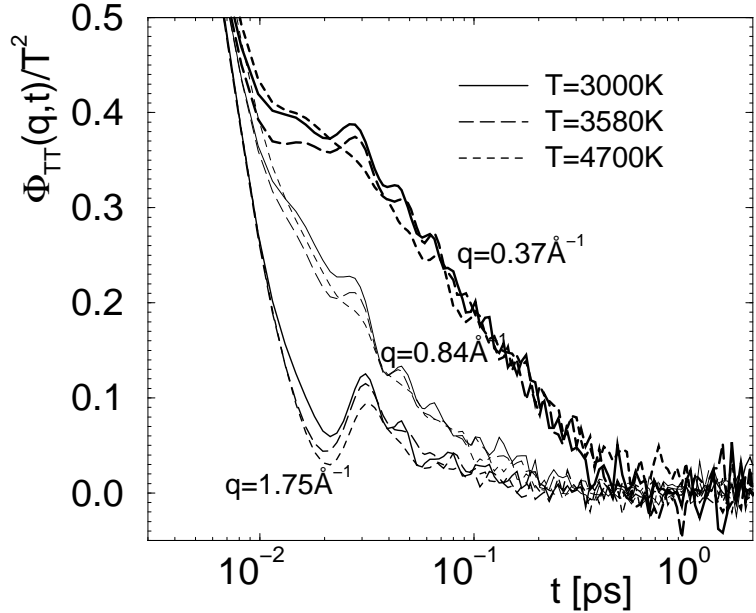,width=10cm,height=8cm} 
\vspace*{2mm}

\caption{Time dependence of the autocorrelation function of the
generalized temperature fluctuations $\delta T_q(t)$. The different 
line styles correspond to different temperatures and the different 
thickness to different wave-vectors.
}
\label{fig7}
\end{figure}

The expressions derived in Sec.~\ref{sec2} were valid for all wave-vectors
$q$ and only at the end, i.e. in Eq.~(\ref{eq17}), we restricted ourselves
to $q=0$ in order to obtain the equation relating the frequency dependent
specific heat to the temperature fluctuations. After having investigated
so far the temperature dependence of $c(z)$, we now turn our attention
to $\Phi_{TT}(q,t)$, the autocorrelation function of $\delta T_q(t)$, which
measures the fluctuations in temperature at finite wave-vectors. From
the definition of $\Phi_{TT}(q,t)$ it is clear that this function
should scale like $T^2$. That this is indeed the case is shown in
Fig.~\ref{fig7}, where we show $\Phi_{TT}(q,t)/T^2$ for various wave-vectors
and temperatures. Since for each wave-vector the curves for the different
temperatures fall on top of each other we recognize immediately that
the $T^{-2}$ dependence is correct. Note that this weak temperature
dependence for $q>0$ is in strong contrast to the one found for $q=0$
(see Fig.~\ref{fig1}b). It reflects the fact that the thermal
conductivity $\lambda$ is only a weak function of temperature, see
Eq.~(\ref{eq23}).

\begin{figure}[h]
\psfig{figure=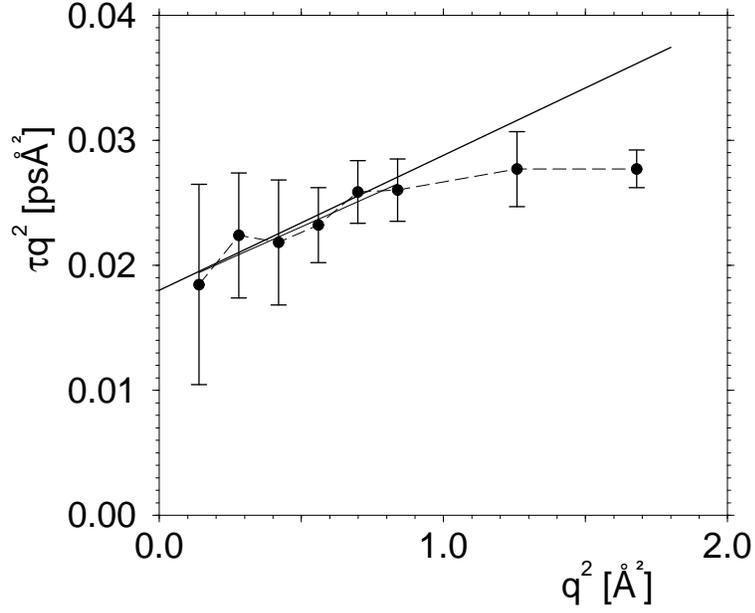,width=10cm,height=8cm} 
\vspace*{2mm}

\caption{Relaxation time for the autocorrelation function $\Phi_{TT}(q,t)/T^2$
multiplied by $q^2$ versus the square of the wave-vector. The
straight line is a fit with the functional form given by
Eq.~(\protect\ref{eq23}). This line is given by 
$q^2 \tau = 0.0180$ps\AA$^{-2}+0.0101q^2$ps.
}
\label{fig8}
\end{figure}

From the plot we also see that the typical time scale over which the
correlation functions decay, increases with decreasing wave-vector,
in qualitative agreement with Eq.~(\ref{eq23}) which predicts a $q^{-2}$
dependence. To determine the $q$-dependence of this decay we define
a decay time $\tau(q)$ as the time it takes the correlation function
to decay to 0.1 of their initial value. The wave-vector dependence
of $\tau(q)$ is shown in Fig.~\ref{fig8} where we plot $q^2 \tau(q)$
versus $q^2$.  (Since within the accuracy of our data the temperature
dependence of $\Phi_{TT}(q,t)$ is independent of $T$ we show only one
set of data points.). From this figure we recognize that for small
wave-vectors the relaxation time scales indeed like $q^{-2}+const$, as
expected from hydrodynamics (straight line), that however, this linear
dependence breaks down for large wave-vectors. Furthermore we see that the
slope of this straight line is positive, which means that the second term
in Eq.~(\ref{eq23}) is larger than zero. From this equation it follows
that the intercept of the straight line with the abscissa is given by
$n c_p^{eq} /\lambda$, where $n$ is the particle density and $\lambda$
is the thermal conductivity. We read off an intercept 0.0180 ps/\AA$^2$
and with the specific heat of 4$k_B$ per particle, see Fig.~\ref{fig3}a,
we obtain $\lambda=2.4$W/Km. This value is in good agreement to the one
determined by a completely different method in the simulation by Jund and
Jullien who found $\lambda\approx 1.3$W/Km around 1000K~\cite{jund99}. The
experimental values for this quantity range between 2 and 3W/Km at high
temperatures if $T\geq 1000 K$~\cite{mazurin83}, i.e. our value is in
agreement also with the experimental data. (Note that in experiments
it is found that the $\lambda$ is a strong function of temperature for
temperatures below $\approx 1000$K. For higher temperatures it seems,
however, to level off and thus it can be extrapolated reasonably safely to
temperatures in the melt. Since this temperature dependence of $\lambda$
is due to anharmonicities one can conclude that these become effectively
independent of $T$ in the temperature range of the supercooled melt.)

\section{Summary and Discussion}
\label{sec5}
The goal of this paper is to show how the frequency dependent specific
heat, $c_v(z)$, is related to the dynamics of the particles on the
microscopic level.  For this we use the Mori-Zwanzig projection operator
formalism to derive an exact expression for $c_v(z)$ (Eq.~(\ref{eq16})).
This expression allows us to identify the physical mechanism which
causes the frequency dependence of $c_v(z)$, namely the relaxation
of the potential energy during the structural relaxation.  Using an
exact transformation formula by Lebowitz {\it et al.}, we obtain an
equation which relates $c_v(z)$ to the Laplace transform of $K(t)$, the
autocorrelation function of temperature fluctuations (Eq.~(\ref{eq21}))
in the microcanonical ensemble, and which thus can be used to determine
$c_v(z)$ from a computer simulation.  This relation has been derived
previously by Nielsen~\cite{nielsen99} on the basis of thermodynamic
arguments. In contrast to that approach we are, however, also able to
generalize the correlator $K(t)$ to finite wave-vectors and to relate
the time dependence of these quantities to the thermal conductivity of
the system, Eq.~(\ref{eq23}).

By using molecular dynamics computer simulations of a simple but quite
realistic model for silica, we have determined the time and temperature
dependence of $K(t)$. We see that at low temperatures this correlator
shows a two-step decay, similar to the one that is found in the time
correlation functions for structural quantities, such as the intermediate
scattering function. In contrast to these correlators the height of the
plateau at intermediate times {\it decreases} with decreasing temperature,
a trend that can be understood by realizing that at very low temperatures
$K(t)$ is directly related to the autocorrelation function of the velocity.

From the time dependence of $K(t)$ we have calculated the frequency
dependent specific heat. In contrast to previous numerical investigations
the accuracy of our data is high enough to analyze in detail the frequency
dependence of the real and imaginary part of $c(z)$. We find that at very
high frequencies the value of $c'(\omega)$ is the one of an ideal gas and
that with decreasing frequency it shows a rapid increase which is due to
the vibrational excitations of the system. At low temperatures we see that
$c'(\omega)$ shows a second increase at frequencies which correspond to
the time scales of the structural relaxation. This frequency dependence
is also reflected in $c''(\omega)$ where the first and second increase
are reflected by the microscopic and $\alpha-$peak, respectively. Thus
we find that the frequency dependence of $c(z)$ is qualitatively very
similar to the one found in the dynamical susceptibility for structural
quantities, which shows how intimately connected these quantities
are. Further evidence for this can be obtained from the observation
that the location of the $\alpha-$peak in $c''(\omega)$ shows the same
temperature dependence as the structural relaxation, in agreement with
experimental findings~\cite{menon96,leyser95,jeong95}.

From the height of the two mentioned steps in $c'(\omega)$, we are
able to determine the vibrational and the configurational part of the
specific heat. We find that the former is significantly higher than the
latter, which is in agreement with the experimental observation that in
strong liquids the drop in the specific heat at the glass transition is
relatively small.

Finally we have calculated the time dependence of the autocorrelation
functions of temperature fluctuations at finite wave-vectors. In agreement
with our theoretical prediction, these functions decay much faster than
the one for $q=0$, i.e. $K(t)$, and depend only very weakly on temperature. 
From the $q-$dependence of the relaxation
time of this correlator we calculate the thermal conductivity $\lambda$
and find good agreement of our value with the one in experiments and
a computer simulation in which $\lambda$ was determined by a different
method.

We also point out that, since our simulations have been done at
constant volume, it is clear that the frequency dependence of $c(z)$
and the strong temperature dependence of $\omega_{max}$, the location
of the $\alpha-$peak in $c''(\omega)$, is {\it not} the result of
the frequency and temperature dependence of the macroscopic density.
Some time ago Zwanzig proposed (essentially) the following mechanism
for the $T-$dependence of $\omega_{max}$~\cite{zwanzig88}: A change in
temperature will in general give rise to a change in density (since most
real experiments are done at constant pressure and not density). Due to
the high value of the bulk viscosity, this volume relaxation will be slow
and occur on the time scale of the $\alpha-$relaxation, and hence $c_p$
will be frequency dependent. Since in turn the frequency dependence
of the viscosity is due to the slow relaxation of the structure on
the microscopic scale, Zwanzig thus argued that the reason for the
$T-$dependence of $\omega_{max}$ is just an {\it indirect} effect of the
slow microscopic dynamics. Since in a system with constant volume this
mechanism is not present and our simulations demonstrate that $\omega_{max}$
does show a strong $T-$dependence, we conclude that the reason for this
dependence must be a different one.

Finally we also mention that from the shape of the $\alpha-$peak in
$c''(\omega)$ it is also possible to calculate the time dependence
of the enthalpy in a cooling and heating experiment. For this we made
simulations in which the sample was first cooled with a finite cooling
rate through the glass transition temperature and subsequently reheated
above $T_g$. Using the {\it equilibrium} data for $c(z)$ we were
able to reproduce accurately the time and temperature dependence of the
enthalpy~\cite{scheidler_dipl,scheidler00b}, which shows that if one
knows the {\it equilibrium} quantity $c(z)$, one is also able to
predict the system in an out-of-equilibrium situation.

Acknowledgments: We thank U. Fotheringham and F. Sciortino for useful
discussions. This work was supported by Schott Glas and the DFG
under SFB~262. Part of this work was done at the HLRZ in J\"ulich.

\begin{appendix}
\section{Relation between the density of states and the autocorrelation
function of the kinetic energy for a harmonic system}

For a purely harmonic system with the Hamiltonian
\begin{equation}
H = \frac{{\bf p}^2}{2 m} + \frac{m \Omega^2{\bf r}^2}{2}
\label{eqa1}
\end{equation}
momenta and space coordinates are Gaussian
variables. This simplifies considerably the evaluation of the
normalized autocorrelation function of the kinetic energy
$K(t)$. For a set of Gaussian variables with zero mean the four
point correlation function $\langle ABCD\rangle$ can be expressed by
the two point correlation functions:
\begin{equation}
\langle ABCD\rangle = \langle AB \rangle 
\langle CD\rangle  + \langle AC\rangle 
\langle BD\rangle  + \langle AD\rangle
\langle BC\rangle .
\label{eqa2}
\end{equation}
Using this relation, the autocorrelation function $\langle p_\mu^2(t)
p_\nu^2 \rangle$ can be written as
\begin{equation}
\langle p_\mu^2(t)p_\nu^2 \rangle = 2 \langle p_\mu(t) p_\nu \rangle^2
+ \langle p_\mu^2 \rangle \langle p_\nu ^2 \rangle \quad .
\end{equation}  
Since $\langle p_\mu(t) p_\nu \rangle = \delta_{\mu \nu} \langle
p_\mu^2 \rangle \cos \Omega t $, the autocorrelation function
$\Phi_{TT}(t)$ is given by
\begin{eqnarray} 
\Phi_{TT}(t) &=& \frac{1}{9 m^2 k_B^2} (\langle {\bf p}^2(t) {\bf
p}^2 \rangle - \langle {\bf p}^2\rangle ^2) \\
&=& \frac{2}{9 m^2 k_B^2}  \sum_\mu \langle p_\mu(t) p_\mu \rangle^2 
\\
&=& \frac{T^2}{3} (\cos( 2 \Omega t) + 1)  \quad .
\label{eqa3}
\end{eqnarray}
All averages are in the canonical ensemble and we used that $\langle
p_\mu^2 \rangle = k_B T$.  From Eq. (\ref{eq19}) we know, by using the
value of the specific heat of a harmonic oscillator in three dimensions
$c_v^{eq} = 3 k_B$, that $\Phi_{TT}^{(mc)}(t) = \Phi_{TT}(t) - T^2/3$.
By defining the velocity autocorrelation function $J(t) = m \langle
{\bf v}(t) {\bf v} \rangle$  and noting, that for a harmonic oscillator
$(\vec{v}(t)| \vec{v}) = 3 k_B T /m \cos(\Omega t)$  we obtain the result
\begin{equation}
K(t) = \frac{J(2t)}{6 k_B T} \quad .
\label{eqa4}
\end{equation}
Taking the Fourier transform of (\ref{eqa4}) and using, that the density
of states $g(\omega) = 2 J(\omega)/3 k_B T$~\cite{dove93}, we arrive at the result
\begin{equation}
\hat{K}(\omega) = \frac{g(\omega/2)}{8} \quad .
\label{eqa5}
\end{equation}

\end{appendix}

\begin{references}
\bibitem{footnote1}
Note that this definition of $T_g$ is not the same as the one often
used and which is defined by requiring that at $T_g$ the viscosity of
the system is $10^{13}$ Poise.
\bibitem{footnote6}
In Eq.~(\protect\ref{eq14}) we made use of the relation $Q_{\rho} E_q^P\equiv
Q_{\rho} Q E_q^P\equiv Q E_{\rho}^P$.
\bibitem{footnote}
Note that in equilibrium the specific heat at constant volume and
constant pressure are related by $c_p-c_v=TV\alpha^2/\kappa_T n$,
where $\alpha$ is the thermal expansion coefficient and $\kappa_T$
is the isothermal compressibility. For the frequency dependent case this
relation is generalized to $c_p(z)-c_v(z)=T \alpha^2(z) K_B(z)/N$, where
$\alpha(z)=V^{-1} \partial V(z)/\partial T(z)$ is the frequency 
dependent thermal expansion coefficient~\protect\cite{gotze89,latz91}.
\bibitem{footnote4}
Note that the finite width of the peak at $\omega=0$ as shown in the figure
is due to our finite resolution.
\bibitem{footnote2} 
Note that at intermediate and high temperatures the $\alpha-$peak and the
microscopic process are not separated well from each other. Thus in
order to determine the height of the low-frequency step in $c'(\omega)$
it is necessary to extrapolate this step to higher frequencies. This
was done by means of the sum of two Kohlrausch-Williams-Watts laws,
which at this point should be considered as mere fitting functions
without physical significance. More details on this can be found in
Ref.~\protect\cite{scheidler_dipl}.
\bibitem{footnote3}
Note that the experimental values are for $c_p$ whereas the theoretical
ones are for $c_v$. However since for silica the thermal expansion
coefficient $\alpha$ is very small\protect\cite{mazurin83}, the
difference between $c_p$ and $c_v$ is also small, as can be seen from
the relation $c_p-c_v=T\alpha^2/\kappa_T n$, where $\kappa_T$ is the
isothermal compressibility.
\end{references}
\end{document}